\documentclass[]{rsos}
\PassOptionsToPackage{hyphens}{url}\usepackage{hyperref}
\usepackage{hyperref}
\usepackage{lineno}

\usepackage{natbib}
\begin{document}

\title{Percolation Transition of \textit{K}-Destinations Reveals Universal Backbone of Urban Mobility Networks}

\author{
Weiyu Zhang$^{1,\dag}$, Furong Jia$^{1,\dag}$, Jianying Wang$^{2}$, Yu Liu$^{1}$ and Gezhi Xiu$^{1}$}

\address{$^{1}$Institute of Remote Sensing and Geographical Information Systems, Peking University, Yiheyuan Road 5, Haidian District, Beijing, 100871, China \\ 
$^2$Thrust of Urban Governance and Design, Society Hub, The Hong Kong University of Science and Technology (Guangzhou), Guangzhou, China \\
\dag These authors contributed equally to this work.}

\subject{urban science, complexity science}


\corres{Jianying Wang\\
\email{jianyingwang@hkust-gz.edu.cn}\\
Gezhi Xiu\\
\email{gezhixiu@gmail.com}
}

\begin{abstract}
Cities are held together by a small subset of recurrent origin--destination ties, but extracting this backbone from noisy mobility data remains difficult, and most models still prioritize distance over connectivity. We build directed mobility networks from 48 months of mobile-phone data across eight U.S. cities (2018--2021) and apply a rank-based percolation filter that retains, for each origin, its top-$K$ destinations. Defining $K^*$ as the smallest cutoff that yields a strongly connected network, we find that city-specific thresholds cluster tightly across space and time, with a representative value near $K\approx130$. Degree-preserving and gravity-style baselines do not recover this scale, underscoring the importance of selectively preserved long-range functional ties. We then define $\mathrm{PPD}(K)$, an origin-level mobility concentration measure, and show that socioeconomic associations peak around the percolation-derived integration regime. Together, these findings offer a principled method to extract a minimal mobility backbone and connect its structure to persistent socioeconomic variation.

\end{abstract}

\begin{fmtext}
\linebreak
\end{fmtext}
\maketitle

\section{Introduction}
\noindent
Human mobility offers a direct lens on how cities functionally connect, but relating travel flows to urban structure requires separating persistent collective regularities from individual behaviours~\cite{pflieger2010introduction, bettencourt2007growth, bettencourt2021complex, sevtsuk2012urban, barrat2004architecture, hashemi2016weight, schlosser2020covid, serrano2009extracting}. Although individual travel behaviours show regularities with stable activity rates and heavy-tailed destination frequencies~\cite{alessandretti2018evidence, alessandretti2020scales, schlapfer2021universal}, interaction models usually do not keep such preferences, and still treat space primarily through proximity and distance decay~\cite{ren2014predicting, stouffer1940intervening, simini2012universal}. Yet many of the ties that integrate a city are not strictly local: residential areas repeatedly connect to a limited set of shared activity anchors, often beyond their immediate surroundings. This motivates the central question of this work: do collective travel choices from residential areas---as aggregates of individuals with similar latent demographic and socioeconomic characteristics~\cite{singleton2020mapping, singleton2024segmentation, singleton2022estimating}---reveal a stable backbone of preference-driven origin--destination connections that cannot be reduced to proximity alone?

Capturing collective preferences in place-to-place interactions requires a shift in perspective. The literature on place-to-place interactions emphasizes destinations' attractiveness~\cite{schlapfer2021universal, pappalardo2015returners}, which inherits diffusion-like mobility patterns with distance decay. We thus treat residential areas as the primary units of origin and ask how strongly residents of a common origin concentrate their outflows into a shared set of destinations. These recurrent destinations define functional links that shape neighborhood roles and city-wide integration~\cite{pappalardo2015returners}. To further mitigate distance decay effects, we adopt an explicitly network-based view to move beyond proximity as the organizing principle: the ``functional neighborhood'' of an origin is defined by connectivity—who it stably reaches—rather than by adjacency. This view is consistent with evidence that individual mobility is dominated by a relatively small, stable activity set (on the order of tens of locations)~\cite{alessandretti2018evidence, alessandretti2020scales}.


To infer this topology and separate its structurally significant core from a background of weak ties, we model the city as a directed spatial interaction network from residential origins to destinations and progressively filter it by retention rank~\cite{shante1971introduction, li2021percolation, li2015percolation}. Specifically, for each origin we retain only its top \(K\) most frequently visited destinations and track how global connectivity towards the percolation regime evolves as \(K\) increases. In our data, increasing \(K\) does not primarily cause a few large components to merge; rather, a dominant giant component emerges early and then \emph{gradually absorbs} many small cliques and isolated origins as additional strong ties are added. We therefore define a critical cutoff \(K^*\) operationally as the point at which even the least-connected origins become integrated into the giant component (equivalently, where the remaining disconnected mass becomes negligible). The subnetwork at \(K^*\) provides a principled, minimal approximation of the city’s mobility backbone: it retains only the strongest origin--destination ties while achieving near-complete functional integration.


We empirically evaluate this framework using monthly aggregated mobile phone mobility data across eight major U.S. cities over 48 months~\cite{safegraph_patterns_2022}. Across cities and over time, we find that the estimated critical cutoff \(K_i^*\) lies within a narrow band, enabling a single representative value (approximately \(K\approx 130\)) to serve as a consistent cross-city reference point for comparison. The filtered networks near \(K^*\) also display stable, heavy-tailed connectivity patterns, indicating that a small subset of destinations function as recurrent hubs in the backbone.

Finally, we connect this structural backbone to socioeconomic heterogeneity. For each origin, we define the \emph{Proportion of Principal Destinations} \(\mathrm{PPD}(K)\) as the share of its outflow trips captured by its top \(K\) destinations, and we examine how the association between \(\mathrm{PPD}(K)\) and socioeconomic indicators varies with \(K\). We find that these associations are strongest in the percolation regime around \(K^*\): networks with \(K\ll K^*\) are too sparse to reflect integrated urban function, while networks with \(K\gg K^*\) increasingly incorporate weak ties that dilute signal. In this sense, percolation filtering acts as a principled spatial interaction ``denoiser'', identifying a structurally grounded scale at which mobility concentration most clearly aligns with socioeconomic variation. We emphasize that these results establish robust associations (and predictive structure) rather than a complete causal explanation, since trip frequencies alone do not encode the composition or quality of opportunities at destinations. Together, our findings provide a transferable framework for extracting a stable mobility backbone that links urban connectivity to persistent socioeconomic patterning.

\section{A model to extract the urban mobility backbone through percolation}

Intracity travel combines routine, necessity-driven movements (\emph{exploitation}, e.g., commuting and errands) with occasional visits to less familiar destinations (\emph{exploration}, e.g., dining and leisure)~\cite{liu2012understanding, song2010limits, pappalardo2015returners}. Our aim is to extract a sparse, structurally meaningful ``backbone'' of urban mobility that persists beneath this mixture.

For each city and month $t$, we construct a directed, weighted origin--destination matrix $W(t)\in\mathbb{R}_+^{N_c\times N_c}$ over $N_c$ fine-grained spatial units (e.g., U.S.\ Census Block Groups). Each entry $W_{ij}(t)$ is the total number of observed trips from origin $i$ to destination $j$ during month $t$.
Conceptually, we interpret $W(t)$ as the superposition of recurrent structure and residual variability,
\begin{equation}
W(t)=\langle W\rangle(t)+R(t),
\end{equation}
where this decomposition is used to motivate filtering by strength rather than to imply an explicit statistical factorization.


To suppress weak ties and retain recurrent origin--destination structure, we construct a filtered matrix $W^{(K)}(t)$ by keeping only the top-$K$ destinations for each origin. For each origin $i$, let $\mathrm{Top}_K(i,t)$ be the set of $K$ destinations with the largest values of $W_{ij}(t)$ (breaking ties deterministically). We define
\begin{equation}
W^{(K)}_{ij}(t)=
\begin{cases}
W_{ij}(t), & j\in \mathrm{Top}_K(i,t),\\
0, & \text{otherwise.}
\end{cases}
\end{equation}
Thus, $K$ controls how many principal destinations each origin retains. Edge weights are used to rank destinations, but reachability is evaluated on the induced directed graph (weights do not affect paths).

From $W^{(K)}(t)$ we form a directed graph $G^{(K)}(t)$ on the $N_c$ spatial units, with an edge $i\to j$ if $W^{(K)}_{ij}(t)>0$. Let $O(t)=\{i:\sum_j W_{ij}(t)>0\}$ be the set of origins with nonzero outflow in month $t$, and let $\mathrm{GSCC}(K,t)$ denote the giant strongly connected component (largest strongly connected component) of $G^{(K)}(t)$. We define the monthly integration threshold
\begin{equation}
K^*(t)=\min\{K:\; O(t)\subseteq \mathrm{GSCC}(K,t)\},
\end{equation}
i.e., the smallest cutoff such that all origins belong to the GSCC. The backbone network for month $t$ is then defined as the filtered network at $K^*(t)$, denoted $G^{(K^*)}(t)$. As $K$ increases, $G^{(K)}(t)$ becomes denser and the GSCC expands; $K^*(t)$ provides a connectivity-grounded scale at which a sparse set of strong ties suffices for city-wide reachability.

In the next step, we relate backbone structure to socioeconomic heterogeneity by defining an origin-level mobility concentration measure and examining its association with socioeconomic indicators as a function of $K$.

\begin{figure}[htbp]
    \centering
    \includegraphics[width=\linewidth]{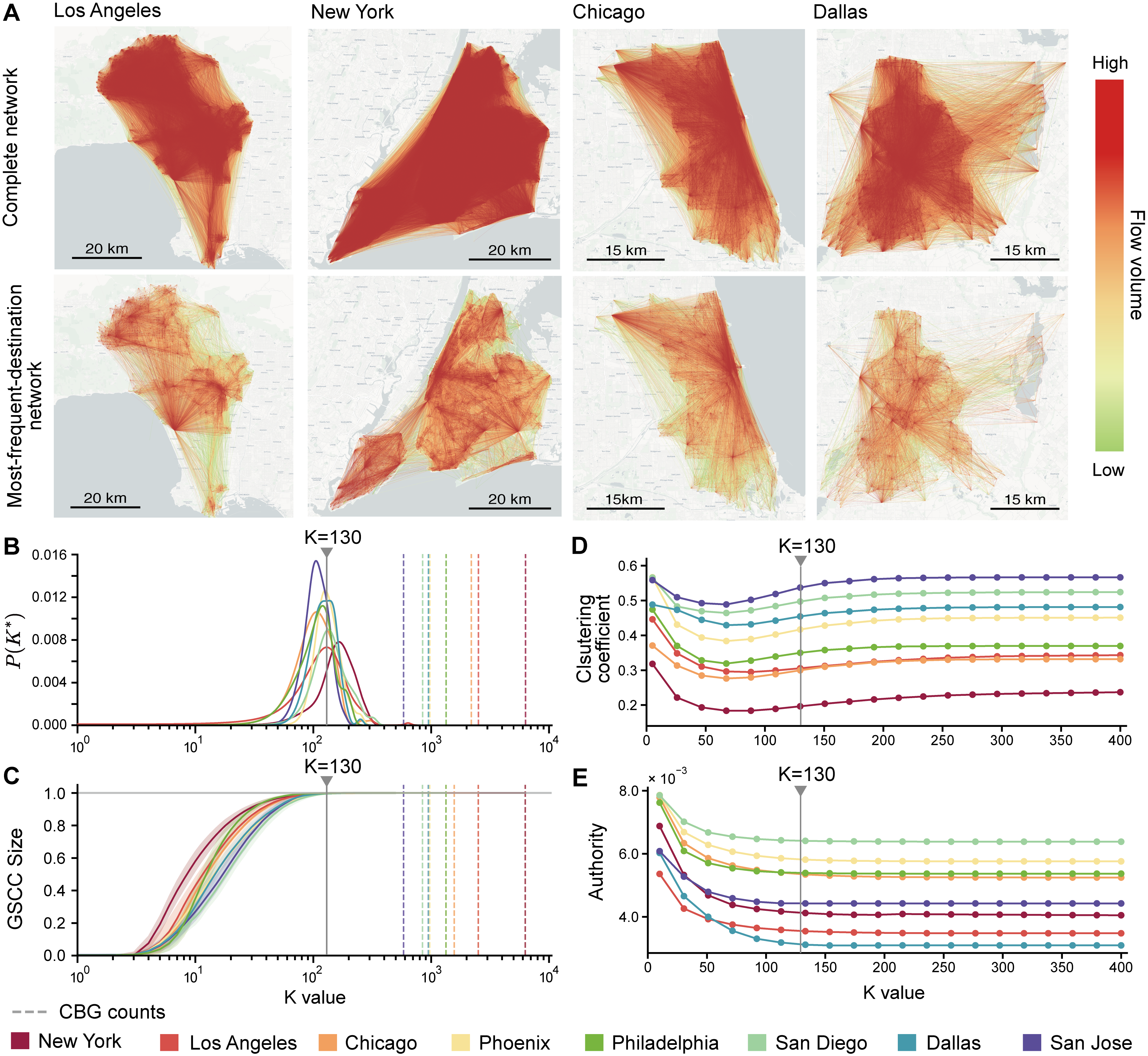}
    \caption{The universal structural conciseness of $K^*$-most-frequent-destination networks for eight U.S. cities. \textbf{A}. Original flow patterns from complete flow networks (top row) and backbone patterns from the $K^*$-most-frequent-destination subnetworks (bottom row). \textbf{B}. Distributions of the city-month integration threshold  $K^*$  and total CBG counts for the eight cities. Each threshold is computed from one monthly mobility network, yielding 48 values per city over January 2018 to December 2021. Distributions are estimated by kernel density estimation (KDE) with bandwidth selected by Scott’s rule~\cite{scott1992multivariate}. The thresholds are concentrated in a narrow range across cities, with a representative scale near  $K \approx 130$, despite substantial variation in city size ranging from 583 in San Jose to 6,493 in New York. \textbf{C}. Fraction of nodes contained in the giant strongly connected component (GSCC) as a function of $K$. Curves show monthly means for each city, and shaded areas indicate $\pm1$ standard deviation across monthly networks. As  $K$  increases, the largest strongly connected component expands continuously and approaches full coverage near the integration regime around  $K \approx 130$. \textbf{D,E}. Topological properties for $K$-most-frequent-destination subnetworks versus the number of most frequent destinations $K$: average clustering coefficient (\textbf{D}) and the mean authority of the top 10 nodes (\textbf{E}).}
    \label{fig:distribution_k}
\end{figure}

\section{Results}

\subsection{Consistency of the integration threshold across cities and over time}

We begin by testing whether the percolation-derived integration threshold \(K^*(t)\) is empirically stable and therefore suitable as a backbone scale for comparison across cities. We compute \(K^*(t)\) for each monthly mobility network (48 months per city) and examine both its distribution and the underlying integration trajectories. Fig.~\ref{fig:distribution_k}A illustrates the filtering effect: the backbone extracted at \(K^*(t)\) preserves the main connective scaffold while suppressing diffuse low-volume links present in the full network. Fig.~\ref{fig:distribution_k}B shows that the resulting city--month thresholds concentrate within a narrow range across all eight cities, despite large differences in the number of CBGs. Fig.~\ref{fig:distribution_k}C complements this result by showing that the GSCC expands smoothly with \(K\) and approaches full coverage in the same integration regime. For cross-city comparison, we use \(K\approx130\) as a representative reference scale within this regime rather than as a city-invariant constant, and the statistical summaries of \(K^*(t)\) for each city are reported in Table S3.

We next examine how the internal structure of the connected core evolves as the network approaches the integration regime. While GSCC size indicates when global reachability is achieved, it does not distinguish whether integration is driven by bridge-like links or by the accumulation of redundant pathways. We therefore compute the average node-wise clustering coefficient (CC) on the \emph{symmetrized} subgraph induced by the GSCC, and report it as a function of \(K\) in Fig.~\ref{fig:distribution_k}D. CC is the fraction of connected triplets that form triangles, and thus quantifies triadic closure within the connected core. For small \(K\), CC decreases as \(K\) increases up to \(\sim100\), consistent with GSCC growth dominated by bridging connections that incorporate new areas without closing many triangles. As \(K\) approaches the integration regime near \(K^*(t)\), CC begins to rise, indicating that additional retained destinations increasingly close triangles and add redundancy among already-integrated nodes. This pattern supports a structural crossover: the network first attains strong connectivity through sparse backbone-like links, and subsequently becomes more triangle-rich as lower-ranked destinations are incorporated.

Finally, we use a hub-centric descriptor to summarize how dominance is distributed among destinations. Fig.~\ref{fig:distribution_k}E reports the mean HITS authority of the ten highest-authority nodes in \(G^{(K)}\)~\cite{kleinberg1999authoritative}. The stabilization of this quantity in the integration regime near \(K^*(t)\) indicates that, beyond the threshold, additional lower-ranked destinations contribute limited changes to the most dominant nodes that support strong connectivity.

Having established that \(K^*(t)\) is stable across cities and time and that the integration regime exhibits consistent structural signatures, we next ask which network ingredients are sufficient to reproduce this integration scale by comparing the empirical networks to randomized and mechanistic baselines.

\subsection{Engineering random networks to probe mechanisms underlying $G^{(K^*)}$}

\begin{figure}[htbp]
    \centering
    \includegraphics[width=0.6\linewidth]{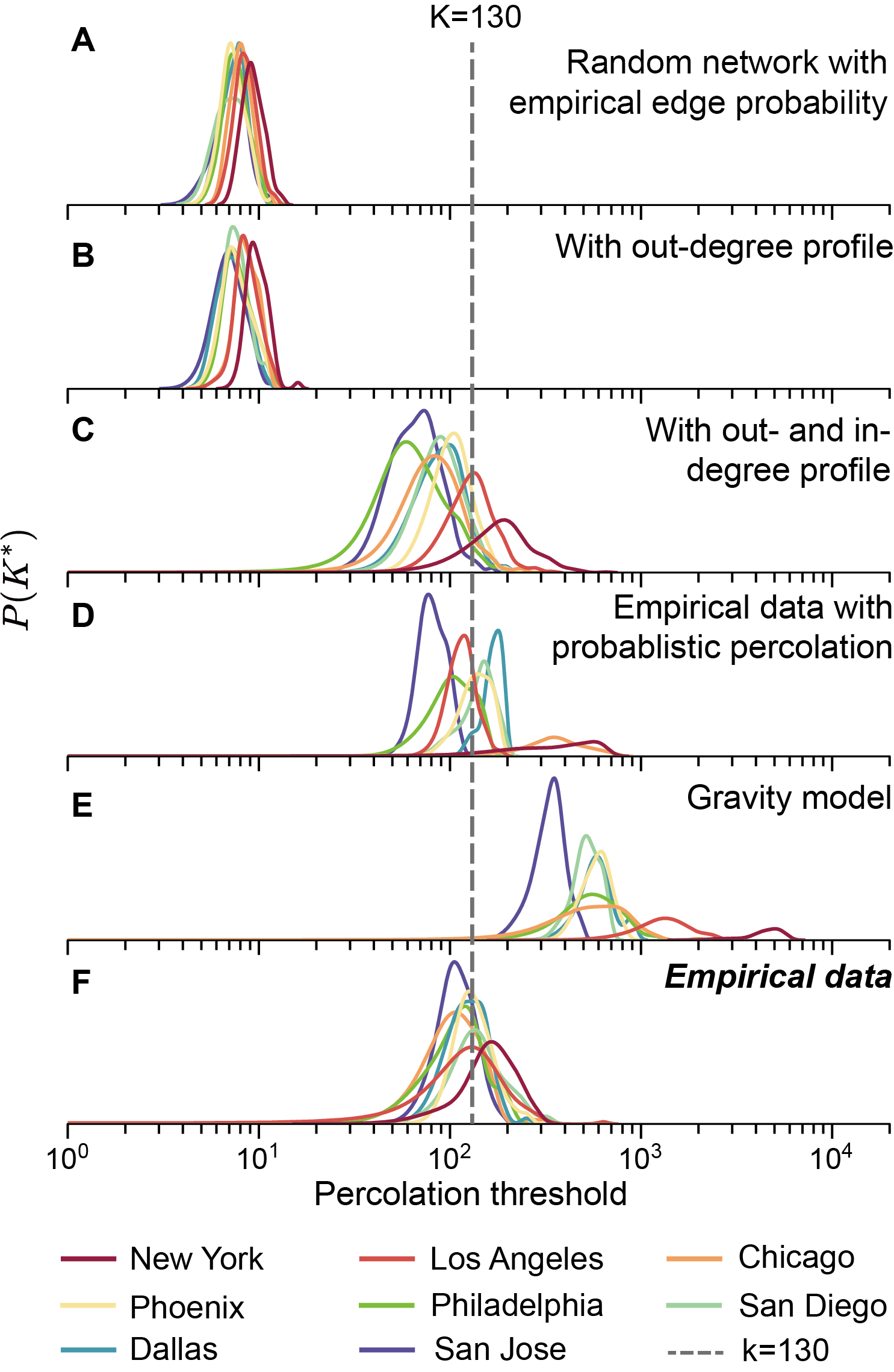}
    \caption{Comparison of integration thresholds between empirical networks and synthetic networks with aligned network properties. \textbf{A-C}. Integration threshold distributions for synthetic random networks preserving different properties of empirical networks: average edge probability (\textbf{A}), out-degree sequence (\textbf{B}), both in- and out-degree sequences (\textbf{C}). \textbf{D}. Integration threshold distributions derived from randomized percolation processes on empirical networks, where destinations are chosen probabilistically rather than deterministically. \textbf{E}. Integration threshold distributions for networks synthesized by the gravity model incorporating spatial distance decay effects. \textbf{F}. Integration threshold distributions for empirical networks. }
    \label{fig:simulation}
\end{figure}

To probe what structural ingredients are responsible for the empirically large integration threshold (with a representative reference scale near \(K\approx130\)), we compare the empirical networks to a sequence of randomized and mechanistic baselines under the \emph{same} top-\(K\) filtering and strong-connectivity criterion (Fig.~\ref{fig:simulation}). For each baseline, we generate synthetic monthly directed networks that preserve selected coarse properties of the empirical data, apply the top-\(K\)-destination filtering, and record the smallest \(K\) at which the filtered graph becomes strongly connected. The baselines are ordered from weak to strong constraints—ranging from simple random wiring to degree-preserving and distance-decay constructions—so that deviations from the empirical threshold can be attributed to missing structural ingredients.

We start from a simple popularity-driven baseline (Fig.~\ref{fig:simulation}A). Each node \(m\) is assigned an origin-specific out-edge propensity \(p_m=E_m^{\mathrm{out}}/(n-1)\), where \(E_m^{\mathrm{out}}\) is the empirical out-degree and \(n\) is the number of CBGs. Directed edges \(m\to j\) are then realized independently with probability \(p_m\). This construction matches expected out-degree levels but removes structured destination choice (i.e., convergence onto shared hubs). The resulting networks typically reach strong connectivity at very small cutoffs (\(K^{*'}\sim10\)), far below the empirical range (Fig.~\ref{fig:simulation}A), indicating that origin-level activity levels alone cannot explain the empirical integration scale.


To test whether degree structure alone can account for the empirical integration scale, we construct a directed configuration-style model that approximately preserves both in- and out-degree sequences. We assign \[p_i^{\mathrm{in}}=\frac{d_i^{\mathrm{in}}}{n-1}, \qquad p_i^{\mathrm{out}}=\frac{d_i^{\mathrm{out}}}{n-1},\] and initiate a directed edge \(i\to j\) with probability \[p_{ij}=c\,p_i^{\mathrm{out}}\,p_j^{\mathrm{in}},\] where \(c\) is chosen to match the empirical mean degree. We again assign weights by sampling from the empirical flow matrix. The resulting thresholds (Fig.~\ref{fig:simulation}C) increase substantially relative to the popularity baselines and often approach \(\sim 100\), indicating that in/out-degree constraints capture a nontrivial portion of the connectivity challenge. However, they still do not fully reproduce the empirical clustering of thresholds around \(K\approx130\), suggesting that additional structure---beyond degrees---contributes to the empirical backbone. Notably, larger cities (e.g., New York and Los Angeles) exhibit higher variance in threshold values under this model, consistent with more Poisson-like wiring once only degree constraints are enforced.

We then test whether the empirical threshold is driven purely by selecting the highest-volume edges. For each origin \(i\), let \(f_i=(f_i^1,\dots,f_i^N)^\top\) be its empirical outflow vector. We construct a randomized top-\(K\) procedure that selects destinations sequentially without replacement: at each step \(K\), each origin chooses an unchosen destination with probability proportional to its remaining flow mass. We stop when the resulting directed graph is strongly connected and record the threshold. Compared to deterministic rank selection, this probabilistic construction yields thresholds with larger cross-city spread and higher temporal variance (Fig.~\ref{fig:simulation}D). This indicates that the empirical integration scale reflects a stable and structured subset of connections—especially for peripheral origins—that is not reproduced by randomized selection even when conditioned on similar flow-volume profiles.

Lastly, we probe proximity-driven structure using a gravity model~\cite{wilson1970entropy}. We reweight empirical flows by
\[\tilde{f}_{ij} = c\,\frac{F_i^{\mathrm{in}}F_j^{\mathrm{out}}}{d_{ij}^2},\]
where \(F_i^{\mathrm{in}}\) and \(F_j^{\mathrm{out}}\) are empirical in- and out-strengths and \(d_{ij}\) is the distance between \(i\) and \(j\). We then derive top-\(K\) networks by ranking destinations under \(\tilde{f}_{ij}\). The resulting thresholds (Fig.~\ref{fig:simulation}E) are typically much larger than empirical, indicating that proximity-weighted interactions alone overemphasize local hubs and delay strong connectivity. Intuitively, long-range functional ties—often lower-ranked by distance-based rules—are disproportionately important for integrating peripheral areas into a strongly connected urban system.

\begin{figure*}[ht]
    \centering
    \includegraphics[width=\linewidth]{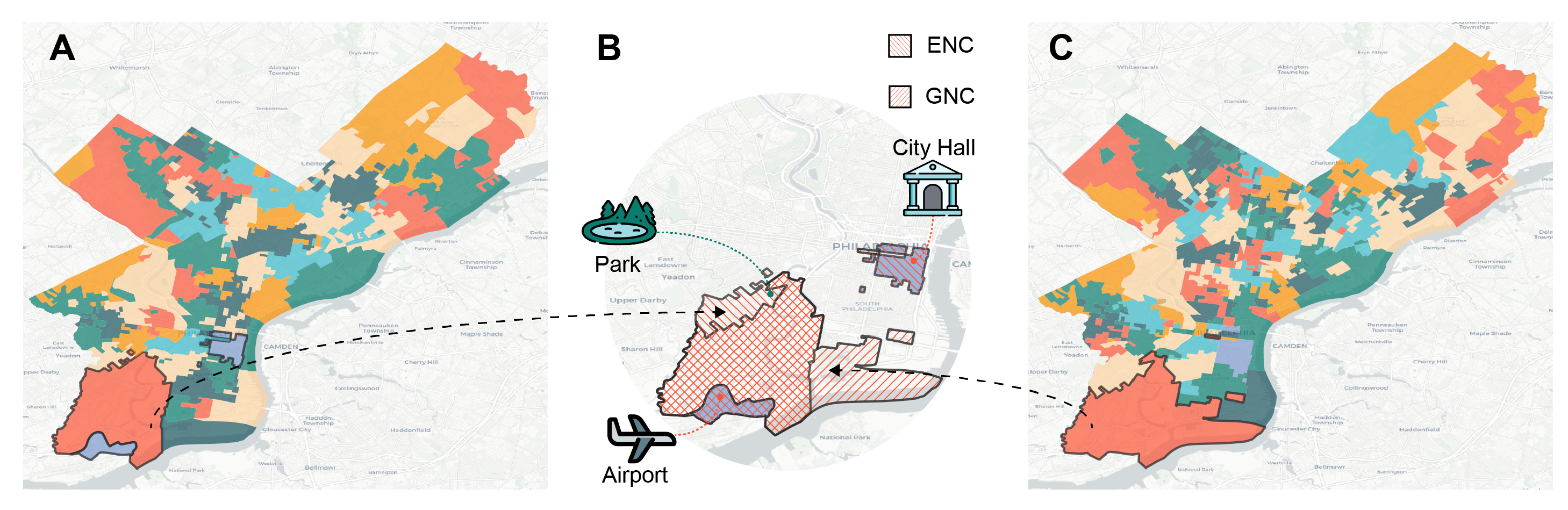}
    \caption{Community detection in Philadelphia using real and simulated networks. \textbf{A}. The community detection results based on the real network, denoted as Empirical Network's Community (ENC). \textbf{C}. The community detection results based on the simulated network, represented by the Gravity Model Network's Community (GNC). \textbf{B}. The best-matching community pair with the highest F1 score.}
    \label{fig:ph_communities}
\end{figure*}

To illustrate how proximity bias alters mesoscopic structure, we apply community detection~\cite{pares2018fluid} to Philadelphia using both the empirical and gravity-style networks at the common reference cutoff \(K=130\). We set the number of communities to 50 and compare partitions using $F1$ overlap between gravity-model communities \(C_j^g\) and empirical communities \(C_i^e\) (Fig.~\ref{fig:ph_communities}). Although the partitions are broadly similar, the mismatches are informative: proximity-based communities tend to group certain high-attractor locations with geographically adjacent areas, whereas the empirical network can place them within communities centred on functionally complementary regions. This qualitative difference is consistent with the gravity baseline underrepresenting long-range functional ties that remain structurally important for strong connectivity.


Taken together, these baselines clarify why the empirical integration threshold concentrates near the \(K\approx130\) regime: neither origin activity levels alone, nor degree constraints alone, nor distance decay alone reproduces the empirical scale. The backbone \(G^{(K^*)}\) instead reflects structured destination choice combined with heterogeneous degrees and selectively preserved long-range ties that collectively stabilize city-wide strong connectivity.

\subsection{Inflow degree distribution at criticality}

To further understand the process by which strong connectivity emerges as \(K\) increases, we examine the evolution of destination in-degrees during the integration of nodes into the GSCC. At the percolation-derived threshold \(K^*(t)\) (Fig.~\ref{fig:distribution_k}B), the top-\(K\) filtering yields the smallest directed subnetwork that is strongly connected. Because \(K^*\) is far smaller than the number of spatial units in a city (\(\sim10^3\) CBGs), strong connectivity must be achieved through a highly selective subset of destinations that are repeatedly shared across many origins. This motivates characterizing the in-degree structure around the integration regime.

\begin{figure}[htbp]
    \centering
    \includegraphics[width=\linewidth]{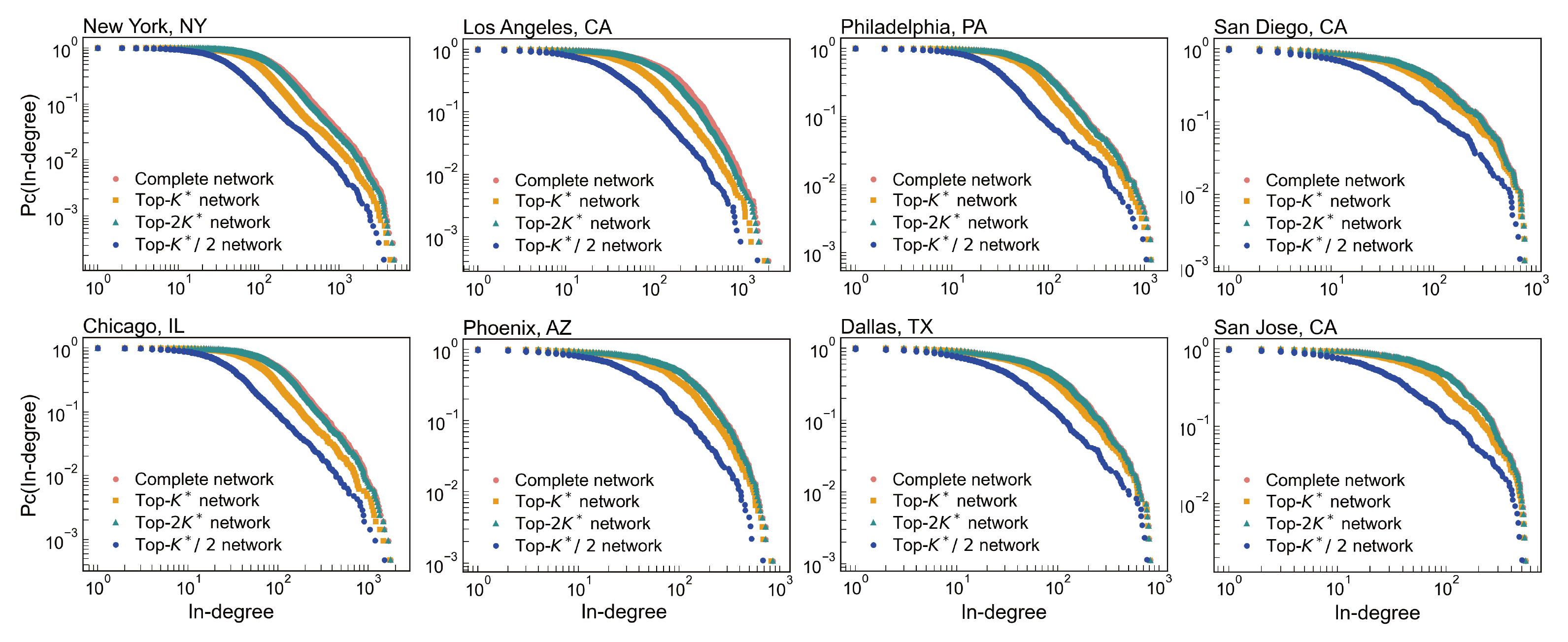}
    \caption{In-degree complementary cumulative distribution functions (CCDFs) for (i) the complete network and (ii) top-\(K\)-destination subnetworks at three connectivity levels: \(K^*(t)\) (city- and month-specific strong-connectivity threshold), \(K^*(t)/2\), and \(2K^*(t)\).}
    \label{fig:powerlaw_fit}
\end{figure}

Fig.~\ref{fig:powerlaw_fit} reports the CCDFs of destination in-degrees at \(K^*(t)/2\), \(K^*(t)\), and \(2K^*(t)\) across the eight cities. Two features are consistent across cities. First, the intermediate-degree range is well approximated by a heavy-tailed form, as indicated by an approximately linear segment on log--log axes. Second, the extreme tail bends downward relative to this intermediate-range scaling: the number of very high-degree destinations is systematically \emph{smaller} than would be predicted by extrapolating the intermediate-range trend. This tail suppression is visible for all cities, indicating that the backbone is hub-dominated but not dominated by an abundance of extremely high-degree hubs.

The way the tail changes with \(K\) clarifies the integration process. Increasing \(K\) from \(K^*(t)/2\) to \(K^*(t)\) steepens the CCDF (\(slope_{K^*}/slope_{K^*/2}=1.1113 \pm 0.0382\)), meaning that additional retained destinations in this window disproportionately expand the low- and mid-inflow population rather than amplifying the very largest hubs. In contrast, expanding from \(K^*(t)\) to \(2K^*(t)\) yields a much smaller slope change (\(slope_{2K^*}/slope_{K^*}=1.0653 \pm 0.0765\)), indicating a diminishing marginal impact on the hub hierarchy once strong connectivity has been achieved. Taken together, these patterns support a structural \emph{saturation} interpretation: the rank window between roughly \(K^*/2\) and \(K^*\) contains many destinations that are critical for integrating remaining nodes into the GSCC, whereas destinations added beyond \(K^*\) increasingly contribute weaker, origin-specific ties that do not substantially reshape the global in-degree structure.

These observations also explain why we subsequently focus on the most connected destinations. Strong connectivity in a filtered directed flow network is often bottlenecked by a small set of highly shared nodes that provide reachability between otherwise weakly coupled parts of the graph. Diagnostics that track the prominence and spatial organization of high-degree destinations therefore provide a direct readout of whether new links primarily reinforce existing hubs or are distributed across a broader set of destinations.

The tail-based interpretation is further corroborated by Supplementary Fig.~S3, which shows a two-stage densification pattern in the mean maximum degree of the GSCC: between \(K^*/2\) and \(K^*\), the growth rate of the most connected node slows, consistent with less concentration of new links onto the top hub. Supplementary Fig.~S4 provides complementary evidence that destinations added between \(K^*\) and \(2K^*\) are spatially less structured (lower Moran's \(I\)) than those added between \(K^*/2\) and \(K^*\), consistent with a shift toward more origin-specific destination additions after the network becomes strongly connected.

\subsection{Socioeconomic composition of frequent destinations}

The percolation-derived backbone provides a compact structural representation of mobility, but it also motivates a functional question: how concentrated is mobility from each origin into its principal destinations, and how does that concentration co-vary with neighborhood socioeconomic characteristics? A cutoff in the integration regime (near the representative reference \(K\approx130\)) is large compared with individual activity sets, yet it remains far smaller than the number of spatial units in a city. This motivates measuring, for each origin, the share of outflows captured by its top destinations and examining how this quantity varies across space and socioeconomic status.

To quantify mobility concentration, we define the \emph{Proportion of Principal Destinations} (PPD) as a function of \(K\). For each origin area \(i\) in month \(t\),
\[
\mathrm{PPD}_i(K,t)=\frac{\sum_{j \in \mathrm{Top}_K(i,t)} W_{ij}(t)}{\sum_j W_{ij}(t)},
\]
i.e., the fraction of outflow trips directed to the origin's top-\(K\) destinations. High \(\mathrm{PPD}(K,t)\) indicates that outflows are concentrated among relatively few destinations, whereas low \(\mathrm{PPD}(K,t)\) indicates a more dispersed destination set. Importantly, \(\mathrm{PPD}\) is a structural measure of concentration only: it does not encode the types, quality, or socioeconomic character of destinations. Two areas with identical \(\mathrm{PPD}\) values could therefore correspond to systematically different opportunity structures.

\begin{figure}
    \centering
    \includegraphics[width=0.6\linewidth]{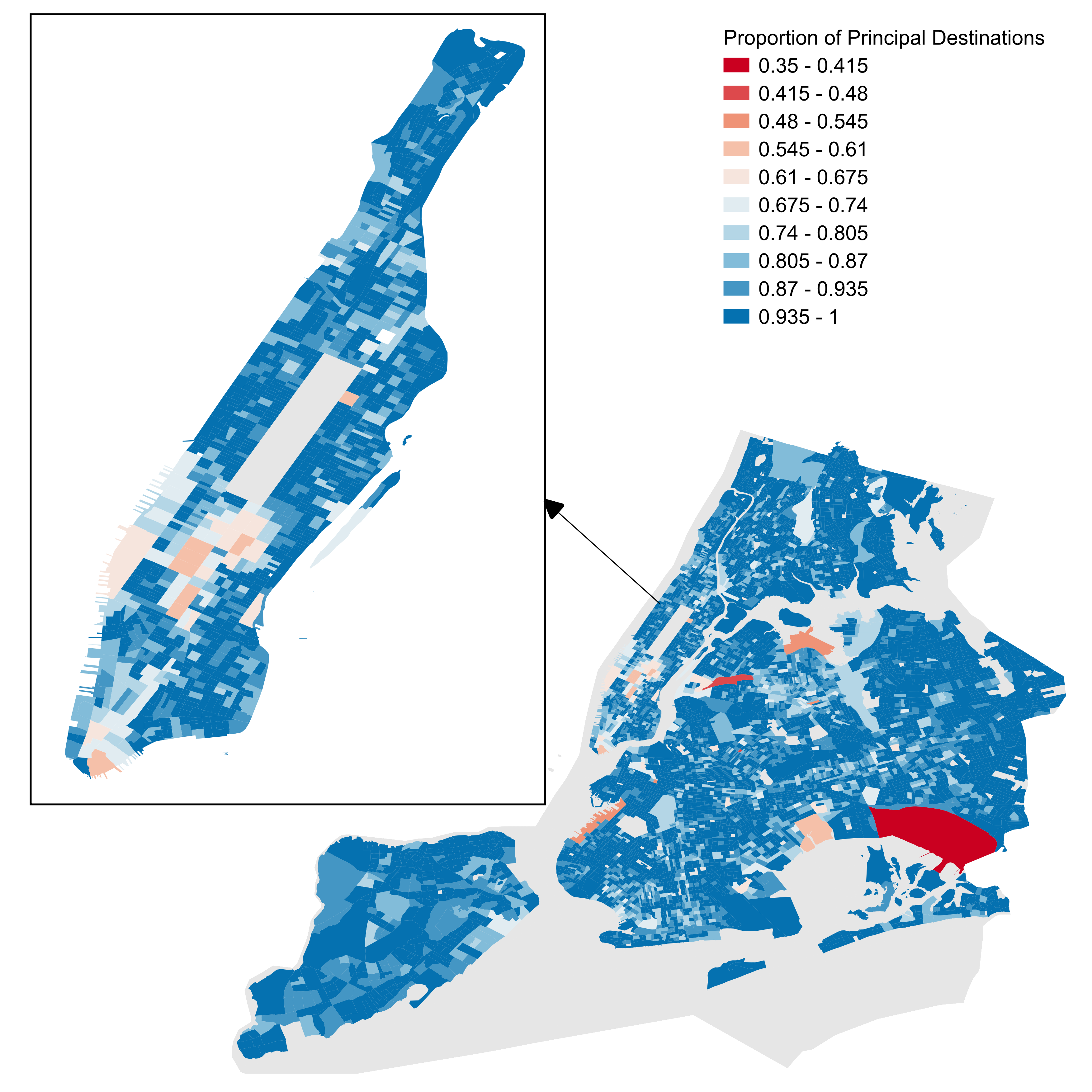}
    \caption{Spatial distribution of the Proportion of Principal Destinations (PPD) in New York City at $K^*$. For a given $K$, PPD is defined as the fraction of trips from each Census Block Group (CBG) directed to its $K$ most-frequented destinations. Areas with high PPD (darker colors) are those where residents' trips are concentrated among relatively few destinations; areas with low PPD (lighter colors) show more dispersed travel patterns. The map reveals spatial heterogeneity in mobility concentration, with low-PPD areas including both non-residential sites (airports, large parks) and several low-income residential neighborhoods. PPD alone does not indicate the quality, accessibility, or socioeconomic character of destinations, and high-PPD areas should not be interpreted as "better" or "more accessible" without additional data on destination characteristics.}
    \label{fig:NYPPD}
\end{figure}

Fig.~\ref{fig:NYPPD} shows the spatial distribution of \(\mathrm{PPD}(K^*)\) in New York. The lowest values frequently occur in CBGs dominated by non-residential public facilities (e.g., major transportation hubs and large parks), reflecting broad catchment patterns rather than neighborhood-level residential behaviour. Among predominantly residential areas, \(\mathrm{PPD}(K^*)\) varies substantially across neighborhoods, highlighting pronounced spatial heterogeneity in mobility concentration. Similar spatial heterogeneity is observed in other cities (see SI).

\begin{figure}
    \centering
    \includegraphics[width=0.9\linewidth]{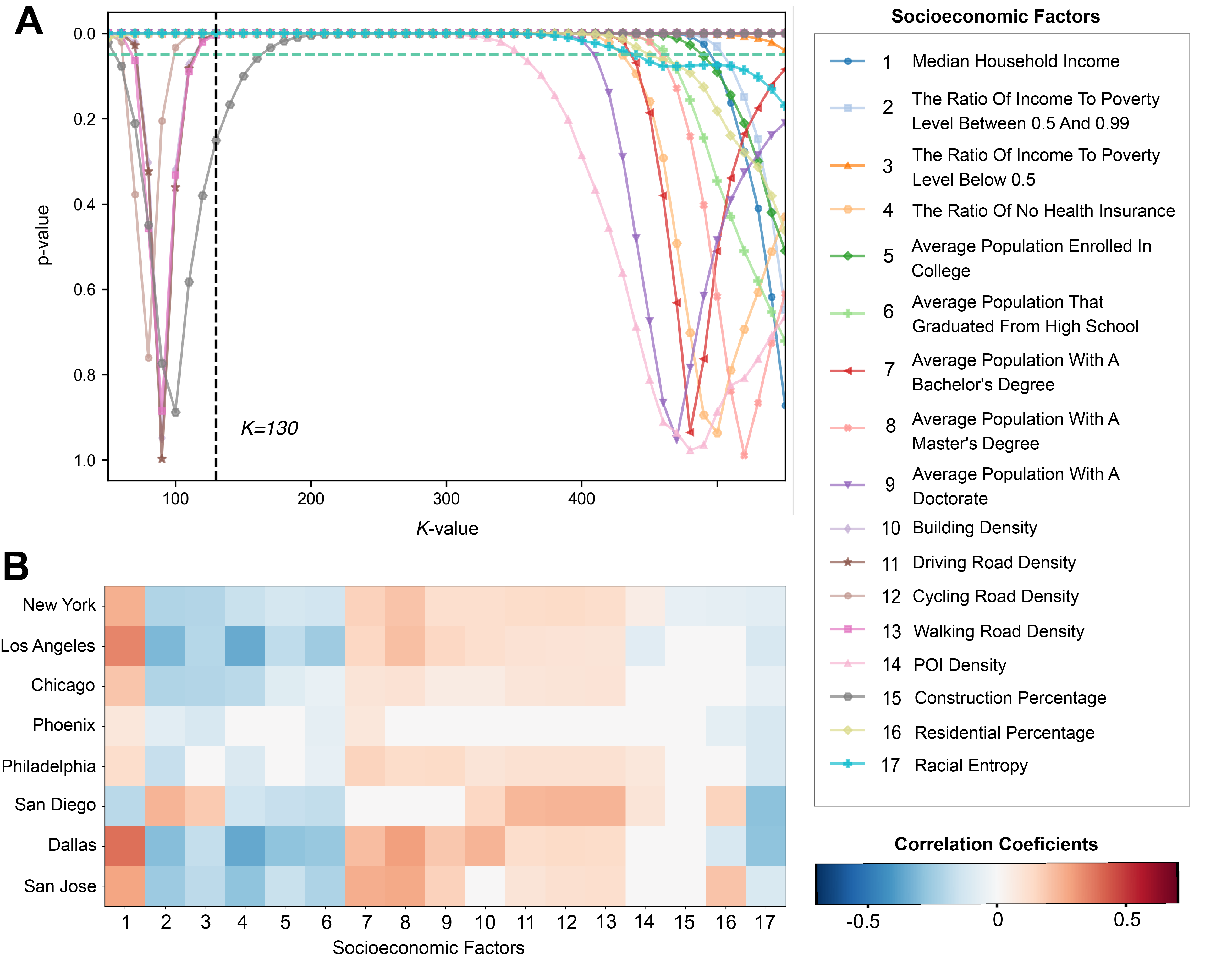}
    \caption{Correlation analysis of socioeconomic factors with the Proportion of Principal Destinations (PPD). \textbf{A}. Statistical significance of correlations between various socioeconomic variables and PPD against the number of frequent destinations (K) in New York City. A significant shift in correlation significance occurs around the representative scale of $K\approx130$, indicating the socioeconomic variables' varying influence on urban mobility patterns at different levels of connectivity. Similar patterns for other cities are shown in Supplementary Fig. S7. \textbf{B}. Correlation analysis between socioeconomic variables and PPD across eight major U.S. cities. Non-significant correlations (P $>$ 0.05) are shown in gray.}
    \label{fig:socio_relationships}
\end{figure}

We quantitatively assess associations between \(\mathrm{PPD}(K^*)\) and socioeconomic characteristics using variables from~\cite{xi2023satellite} and the American Community Survey (ACS)~\cite{USCensusBureau2022}. For each city \(i\) and CBG \(j\), we compute per-capita attributes \(x_{ij}=X_{ij}/P_{ij}\), where \(X_{ij}\) is a raw count and \(P_{ij}\) is total population. We then compute within-city Pearson correlations between \(x_{ij}\) and \(\mathrm{PPD}_{ij}(K^*)\),
\[
r_i=\frac{\sum_{j=1}^{n_i}\left(x_{ij}-\bar{x}_i\right)\left(\mathrm{PPD}_{ij}-\overline{\mathrm{PPD}}_i\right)}
{\sqrt{\sum_{j=1}^{n_i}\left(x_{ij}-\bar{x}_i\right)^2 \sum_{j=1}^{n_i}\left(\mathrm{PPD}_{ij}-\overline{\mathrm{PPD}}_i\right)^2}},
\]
where \(n_i\) is the number of CBGs in city \(i\), and \(\bar{x}_i\) and \(\overline{\mathrm{PPD}}_i\) are city means~\cite{pearson1895vii}. Across cities, we observe systematic associations between mobility concentration and socioeconomic status: median household income correlates positively with \(\mathrm{PPD}(K^*)\) (median \(r\approx0.25\) across cities), while the proportion of population without health insurance correlates negatively (median \(r\approx-0.16\)). Education-related variables also show consistent but heterogeneous associations across measures (Fig.~\ref{fig:socio_relationships}B). We interpret these patterns as statistical co-variation rather than direct causal effects, since flow frequencies alone do not specify the opportunity composition of visited locations.

Finally, we examine how the strength of these associations depends on the cutoff \(K\). Fig.~\ref{fig:socio_relationships}A reports, for New York City, the p-values of the Pearson correlation between \(\mathrm{PPD}(K)\) and selected socioeconomic variables across a range of \(K\). For many variables, the strongest statistical signals occur in the integration regime near the percolation-derived scale. As \(K\) increases beyond this regime, significance levels for many variables decline or plateau. This pattern is consistent with the interpretation that percolation filtering identifies a scale at which mobility concentration is most cleanly related to socioeconomic variation: below this regime the network is under-integrated, while above it additional low-rank destinations contribute increasingly origin-specific ties that reduce interpretability.

\section{Discussion}

We proposed a percolation-based framework to extract a sparse, connectivity-grounded backbone from intracity mobility networks and to relate backbone structure to socioeconomic heterogeneity. The method filters each origin’s outflows by retaining its top-\(K\) destinations and defines a city- and month-specific integration threshold \(K^*(t)\) as the smallest cutoff at which the filtered directed network becomes strongly connected. This yields an operational backbone \(G^{(K^*)}(t)\) determined by network structure rather than by socioeconomic fitting.


Across eight U.S.\ cities over 48 months, the estimated thresholds \(K^*(t)\) concentrate within a narrow range, allowing \(K\approx130\) to serve as a representative cross-city reference scale. We do not interpret 130 as a city-invariant constant; rather, it marks the typical scale of the percolation-derived integration regime. Structural diagnostics further indicate diminishing marginal changes to hub hierarchy beyond this regime, consistent with a saturation effect in which additional lower-ranked destinations contribute increasingly origin-specific ties that are less consequential for global connectivity.


To connect backbone scale to socioeconomic variation, we introduced \(\mathrm{PPD}(K)\), the fraction of an origin’s outflow captured by its top-\(K\) destinations, and evaluated its associations with Census-derived indicators. We observe systematic relationships between \(\mathrm{PPD}(K)\) and socioeconomic measures, with associations often strongest in the connectivity crossover regime near \(K^*(t)\). Substantively, higher-income areas tend to exhibit more concentrated destination use (higher \(\mathrm{PPD}\)), whereas more disadvantaged areas more often exhibit dispersed destination use (lower \(\mathrm{PPD}\)). We interpret these findings as robust statistical co-variation: they document how mobility concentration aligns with neighborhood socioeconomic profiles, but do not identify the causal mechanisms that generate these patterns.

Several limitations clarify the scope of these conclusions. First, mobility flows do not encode the composition or quality of jobs, services, and amenities at destinations; \(\mathrm{PPD}\) is therefore a proxy for mobility concentration rather than a direct measure of opportunity or accessibility. Second, the mobility data are aggregated monthly, which may blur differences between weekdays and weekends, although the stability of \(K^*(t)\) over time suggests the extracted backbone captures stable structure. Third, the socioeconomic results are statistical correlations at the CBG level and do not necessarily reflect individual-level behaviours, and the observed correlations are modest in magnitude. Multiple mechanisms could generate the same association between income and \(\mathrm{PPD}\) (e.g., differences in local amenity availability, employment concentration, or mobility constraints), but \(\mathrm{PPD}\) alone cannot distinguish among them.

Future work could incorporate POI semantics and finer temporal stratification (e.g., weekdays versus weekends) to link backbone ties to opportunity composition and to test candidate mechanisms connecting mobility concentration to inequality. More generally, the percolation framework provides a transferable, model-free method for extracting a minimal mobility backbone and a principled scale for comparing cities. By tying this connectivity-grounded backbone to origin-level mobility concentration, our results offer a compact lens on how stable mobility structure aligns with socioeconomic patterning in urban systems.


\vspace{2em}
\noindent
\textbf{Funding.} This research was funded by the National Natural Science Foundation of China (41830645 and 41971331). G.X. was funded by the China Scholarship Council. J.W. was supported by an RGC Postdoctoral Fellowship awarded by the Research Grants Council of Hong Kong (PDFS2324-4H04).

\noindent
\textbf{Authors’ contributions.} G.X. proposed and designed the research; W.Z., F.J., J.W., and G.X. performed the research; G.X. wrote the paper; Y.L., J.W., F.J., W.Z., and G.X. reviewed the manuscript.

\noindent
\textbf{Competing interests.} The authors declare no competing interests.

\noindent
\textbf{Data accessibility.} All codes are publicly available on GitHub at \url{https://github.com/Mobility-Constant/Top-k-network} and have been archived within the Zenodo repository: \url{https://doi.org/10.5281/zenodo.21131505}.
The socioeconomic data can be accessed from \url{https://github.com/axin1301/Satellite-imagery-dataset} and American Community Survey (\url{https://www.census.gov/programs-surveys/acs}).
The mobility dataset is from SafeGraph Panel Overview data on monthly patterns covering the period from January 2018 to December 2021. The data were publicly available at \url{https://docs.safegraph.com/docs/monthly-patterns} before January 2023 and have since been transferred to Dewey Data, where the dataset can be accessed following the instructions at \url{https://www.deweydata.io/blog/advan-patterns-now-available}.

\noindent

\bibliography{ref_aps}

\end{document}